\documentclass[useAMS,usenatbib,10pt,epsfig]{mn2e}
\usepackage[english]{babel}
\usepackage{amssymb}
\usepackage{graphicx}
\title[ATLANT for Cosmological Recombination]
{Advanced Three Level Approximation for 
Numerical Treatment of Cosmological Recombination}
\author[Kholupenko, Ivanchik, Balashev, and Varshalovich]
{Kholupenko E.E.$^{\rm 1,2}$, Ivanchik A.V.$^{\rm 1,2}$, 
Balashev S.A.$^{\rm 1,2}$, and Varshalovich D.A.$^{\rm 1,2}$\\
$^{\rm 1}$Ioffe Physical-Technical Institute, St.-Petersburg 194021, Russia
\\$^{\rm 2}$ St.-Petersburg State Polytechnical University, Russia}

\begin{document}
\pagerange{\pageref{firstpage}--\pageref{lastpage}} 
\pubyear{2011}

\maketitle

\label{firstpage}
\begin{abstract}
{New public numerical code for fast calculations of the 
cosmological recombination of primordial hydrogen-helium plasma is 
presented. The code is based on the three-level approximation (TLA) model 
of recombination and allows us to take into account some fine 
physical effects of cosmological recombination simultaneously with 
using fudge factors. The code can be found at 
http://www.ioffe.ru/astro/QC/CMBR/atlant/atlant.html
}
\end{abstract}


\section{Introduction}
\hspace{1.1cm}
Cosmological recombination is one of the key processes in the early Universe. 
It determines the epoch of decoupling of radiation from matter and thereby 
determines the epochs at which baryonic matter can start to fall into 
gravitational potential wells created by clustering cold dark matter (CDM). 
Afterwards these CDM+baryonic matter clouds develop into non-relativistic 
gravity bound systems like galaxies (\cite{Peebles65, Peebles68}; 
\cite{Doroshkevich_et_al67}; \cite{Ma_Bertschinger95}). 
The sizes of these proto-objects also 
depend on cosmological recombination via the kinetics of divergence of 
radiation and matter temperatures which determines the critical 
Jeans length. Cosmological recombination affects primordial chemistry 
(\cite{Dalgarno_Lepp87}) 
and correspondingly rate of radiative cooling of collapsing clouds 
(via emission in resonant lines of molecules which depends on the abundances 
of primordial molecules). 

From observational point of view the cosmological recombination is also 
very important process because its kinetics determines the position 
of last scattering surface (\cite{Sunyaev_Zeldovich70}, \cite{Hu_et_al95}). 
This in turn affects the cosmic microwave background radiation (CMBR) 
anisotropy which is 
one of the main sources of information about evolution of the early Universe, 
its composition and other properties. A great number of experiments on CMBR 
anisotropy have been carried out in the last thirty years (Relikt-1 1983, 
COBE 1989, QMAP-Toco 1996, BOOMERanG 1997-2003, MAXIMA 1998-1999, 
WMAP 2001-present day, and many others). Treatment of the results of these 
experiments demands clear understanding of cosmological recombination 
physics. That is why many efforts have been 
made for theoretical investigation of cosmological recombination and 
development of applied numerical codes for modelling of this process 
(e.g. \cite{Zeldovich_et_al68}, \cite{Peebles68}, \cite{Matsuda_et_al69}, 
\cite{Jones_Wyse85}, \cite{Grachev_Dubrovich91}, \cite{Seager_et_al99}). 
Increasing accuracy of experiments in the last decade leads to increase 
of efforts of theorists in the study of details of cosmological recombination 
(see e.g. \cite{Dubrovich_Grachev05}, \cite{Chluba_et_al07, Chluba_et_al10}
\cite{Chluba_Sunyaev09b, Chluba_Sunyaev10b}, \cite{Rubino_Martin_et_al08}, 
\cite{Switzer_Hirata08b}, \cite{Hirata_Switzer08}, \cite{Ali_Haimoud_et_al10}) 
and perfection of numerical codes. 
In the light of soon releases of experimental data from Planck mission 
this task becomes more and more important and urgent (for overview 
of efforts of investigators to find exact recombination scenario 
and to estimate remain uncertainties see e.g. 
\cite{Sunyaev_Chluba09, Shaw_Chluba11}). Today for successful treatment 
of Planck data the minimal required accuracy of numerical codes evaluating 
cosmological recombination is about 0.1\% (on free electron 
fraction) for the epoch of hydrogen recombination and 1\% for the epoch 
of helium recombination. Desirable accuracy is about 0.01\% and 0.1\% 
correspondingly.

The next possible step of the investigations of the early Universe 
in the epochs $z = 800 - 10^{4}$ is the experimental study of CMBR spectral 
distortions originated from cosmological recombination of hydrogen and helium. 
Such experiments would be powerful sources of information about history 
of the Universe in these epochs. 
Thanks to the numerous theoretical works in this field (\cite{Dubrovich75}, 
\cite{Lyubarsky_Sunyaev83, Fahr_Loch91, Rybicki_dell_Antonio93, 
Boschan_Biltzinger98, Dubrovich_Grachev04, Rubino_Martin_et_al08} 
and references therein) one can understand parameters of these 
spectral distortions clearly enough. Of course the 
cosmological recombination is an integral part of modelling of these 
distortions. Note that some simple methods for this problem 
(\cite{Bernshtein_et_al77, Burgin03}) demand the knowledge of derivatives 
of ionization fractions of 
hydrogen and helium, so calculated ionization fractions should be smooth 
as possible (i.e. without sharp numerical features). In spite of the fact 
that detection of CMBR spectral distortions by cosmological recombination 
is impossible today, the rapid progress of measurement equipment for CMBR 
anisotropy allows us to hope that such experiments will be possible in 
the near future. 

Thus the main aim of this work is to present numerical code covering 
investigations of cosmological recombination widely as possible within 
simple model which we used. 
Our code called {\bf atlant} ({\bf a}dvanced {\bf t}hree 
{\bf l}evel {\bf a}pproximation for {\bf n}umerical {\bf t}reatment 
of cosmological recombination) may be useful for regular calculations  
of free electron fraction for treatment of CMBR anisotropy data, 
further investigations of cosmological recombination, theoretical predictions 
of new observational cosmological effects (e.g. \cite{Dubrovich_et_al09, 
Grachev_Dubrovich10}), and comparison 
with other numerical codes (e.g. {\bf recfast} by \cite{Seager_et_al99} 
and \cite{Wong_et_al08}, {\bf RICO} by \cite{Fendt_et_al09}, 
{\bf RecSparse} by \cite{Grin_Hirata10}, 
{\bf HyRec} by \cite{Ali_Haimoud_Hirata10b}, 
{\bf CosmoRec} by \cite{Chluba_Thomas10}). 

\section{Cosmological model}
The standard cosmological model is used. Hubble constant $H$ as 
a function of redshift $z$ is given by the following equation:
\begin{equation}
H(z)= H_0 \sqrt{\Omega_\Lambda+\Omega_K (1+z)^2
+\Omega_m (1+z)^3 + \Omega_{rel}(1+z)^4}
\label{Hubble}
\end{equation}
where $H_0$ is the value of Hubble constant in present epoch, 
$\Omega_{\Lambda}$ is the vacuum-like energy density, $\Omega_K$ is 
the energy density related to the curvature of the Universe, 
$\Omega_m$ is the energy density of non-relativistic matter, which 
includes contributions from CDM and baryonic matter 
$\Omega_m=\Omega_{CDM}+\Omega_{b}$. $\Omega_{rel}$ is the energy 
density of relativistic matter, which includes contributions 
from photons (mainly CMBR) and neutrinos 
$\Omega_{rel}=\Omega_{\gamma}+\Omega_{\nu}$. 

The photon energy density is related to the temperature of CMBR:
\begin{equation}
\Omega_{\gamma}={a_{R}T_{0}^4 \over \rho_{c}c^2}
\end{equation}
where $a_{R}$ is the radiation constant, $T_{0}$ is the 
CMBR temperature at present epoch, $\rho_{c}=3H_{0}^2/\left(8\pi G\right)$ 
is the critical density of the Universe, $c$ is the speed of light, 
$G$ is the gravitational constant.

At the present epoch the relativistic neutrino energy density is related 
with the photon energy density by the following formula:
\begin{equation}
\Omega_{\nu}={7\over 8}N_{\nu}\left({4\over 11}\right)^{4/3}\Omega_{\gamma}
\end{equation}
where $N_{\nu}$ is the effective number of neutrino types.

In this paper (and current version of code) we consider 
$\Omega_{tot}=\Omega_\Lambda+\Omega_K+\Omega_m+\Omega_{rel}=1$, 
so $\Omega_K$ is fixed by relation 
$\Omega_K=1-\left(\Omega_\Lambda+\Omega_m+\Omega_{rel}\right)$.

The total concentrations of hydrogen and 
helium depend on redshift by the following formulas:
\begin{equation}
N_{H}=N_{H0}\left(1+z\right)^3\;,~~~N_{He}=N_{He0}\left(1+z\right)^3
\label{total_concentrations}
\end{equation}
where $N_{H0}$ and $N_{He0}$ are the values of concentrations at present epoch. 
The total concentration of the primordial hydrogen atoms and ions at present 
epoch is given by the following relation:
\begin{equation}
N_{H0}={\rho_{c}\over m_{H}}\Omega_{b}X_{p}
\label{hydrogen_total_conc}
\end{equation}
where $m_{H}$ is the hydrogen atom mass, $X_{p}$ is the primordial 
hydrogen mass fraction (i.e. hydrogen mass fraction after the primordial 
nucleosynthesis).

The total concentration of the primordial helium atoms and ions at 
present epoch is given by the following relation:
\begin{equation}
N_{He0}={\rho_{c}\over m_{He}}\Omega_{b}Y_{p}
\label{helium_total_conc}
\end{equation}
where $m_{He}$ is the helium atom mass, $Y_{p}=\left(1-X_{p}\right)$ 
is the primordial helium mass fraction. Fractions of other elements are 
considered negligible.

The temperature $T$ of equilibrium radiation background depends on redshift 
according to:
\begin{equation}
T=T_{0}\left(1+z\right)
\end{equation}

\section{Hydrogen Recombination}
\subsection{Main equations}
\label{Hydrogen_Main_equations}
The time-dependent behaviour of hydrogen ionization fraction in the isotropic 
homogeneous expanding Universe is described by the following kinetic 
equation (\cite{Zeldovich_et_al68, Peebles68}):
\begin{eqnarray}
\dot x_{HII}=-C_{HI}[\alpha_{HII}\left(T_{m}\right)N_e x_{HII} - 
\beta_{HI}\left(T\right)\exp\left(-{h\nu_{\alpha}\over k_{B}T}\right)x_{HI}]
\label{Zeld_Peebles_eq1}
\end{eqnarray} 
where $x_{HII}=N_{HII}/\left(N_{H}+N_{He}\right)$, $C_{HI}$ is the factor 
by which the ordinary recombination rate is inhibited by the presence of 
HI Ly$\alpha$ resonance-line radiation, $\alpha_{HII}$ is the total 
HII$\rightarrow$HI recombination coefficient to the excited states of HI, 
$N_e=x_{e}N_{H}$ is the free electron concentration ($x_{e}$ is the free 
electron fraction in common notation, see e.g. {\bf recfast}), 
$\beta_{HI}$ is the total 
HI$\rightarrow$HII ionization coefficient from the excited states of 
HI, $\nu_{\alpha}$ is the Ly$\alpha$ transition frequency, $T_{m}$ 
is the kinetic temperature of the electron gas, 
$x_{HI}=N_{HI}/\left(N_{H}+N_{He}\right)$ is the neutral hydrogen fraction. 
Note that ionization 
fractions of ionic components are defined relative to the total number of 
hydrogen and helium atoms and ions $\left(N_{H}+N_{He}\right)$ 
while free electron fraction $x_{e}$ is normalized to the concentration of 
hydrogen atoms and ions, $N_{H}$, as it is accepted commonly 
(so it is necessary for possible use of {\bf atlant} results 
by other numerical codes).

The specific form of inhibition coefficient depends on fine effects 
which are taken into account. In the current version (1.0) of the code 
only radiative feedbacks for resonant transitions are taken into account 
from the whole list of known (considered until now) fine effects. 
Other fine effects are planned to be included in future works. 
So, the inhibition coefficient is given by the formula:
\begin{equation}
C_{HI}={A^{r}_{eff} + A_{2s1s} \over 
\beta_{HI}+A^{r}_{eff} + A_{2s1s}}
\label{C_HI_res}
\end{equation}
where $A^{r}_{eff}$ is the total effective coefficient of 
np$\leftrightarrow$1s transitions, $A_{2s1s}=8.22458$ s$^{-1}$ 
is the coefficient of 2s$\rightarrow$1s two-photon spontaneous 
transition (e.g. \cite{Goldman89}). 

The recombination coefficient $\alpha_{HII}$ is given by the following 
approximation:
\begin{equation}
\alpha_{HII}\left(T\right)=F_{H}
{aT_{4}^{b} \over 1+cT_{4}^{d}}
\label{hydrogen_recomb_coeff}
\end{equation}
where $F_{H}$ is the hydrogen fudge factor by \cite{Seager_et_al99}, 
$T_{4}=T[K]/10^{4}$, and $a=4.309\cdot 10^{-13}$cm$^{3}$s$^{-1}$, 
$b=-0.6166$, $c=0.6703$, $d=0.5300$ are the parameters 
fitted by \cite{Pequignot_et_al91}.

The ionization coefficient $\beta_{HI}$ can be found by using principle of 
detailed balance:
\begin{equation}
\beta_{HI}\left(T\right)=\alpha_{HII}\left(T\right)
g_e\left(T\right)\exp\left({-{h\nu_{c2} \over k_B T}}\right)
\label{hydrogen_ioniz_coeff}
\end{equation}
where $g_e\left(T\right)={(2\pi m_{e} k_{B}T)^{3/2} / h^3}$ is the 
partition function of free electrons, $\nu_{c2}$ is the HI c$\rightarrow$2 
transition frequency (here symbol ``c'' denotes continuum state).

\begin{figure}
\centering
\includegraphics[bb = 20 50 525 525, width=8cm, height=8cm]{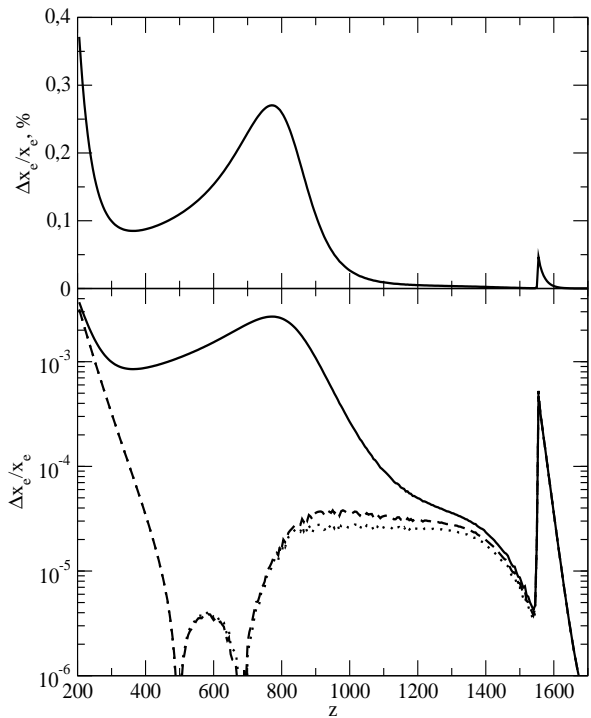}
\caption{
Top panel: the relative difference $\Delta x_{e}/x_{e}$ between free 
electron fractions calculated by {\bf atlant} and {\bf recfast} for 
the period of hydrogen recombination. Bottom panel: 
(i) solid curve corresponds to the relative difference $\Delta x_{e}/x_{e}$ 
between free electron fractions calculated by {\bf atlant} and {\bf recfast} 
[i.e. the same as in the top panel but in logarithmic scale], 
(ii) dashed curve corresponds to the relative difference 
$\Delta x_{e}/x_{e}$ between free electron fractions calculated by {\bf atlant} 
and {\bf recfast} with corrected ionization rate [i.e. ionization coefficient 
$\beta_{HI}\left(T\right)$
and exponential term 
$\exp\left(-{h\nu_{\alpha}/\left[k_{B}T\right]}\right)$ 
in Eq. (\ref{Zeld_Peebles_eq1}) of {\bf recfast} are 
calculated at the temperature of radiation $T$], 
(iii) dotted curve corresponds to the relative difference 
$\Delta x_{e}/x_{e}$ between free electron fractions calculated by 
{\bf atlant} with modified ionization rate [i.e. ionization coefficient 
$\beta_{HI}\left(T_{m}\right)$
and exponential term 
$\exp\left(-{h\nu_{\alpha}/\left[k_{B}T_{m}\right]}\right)$ 
in Eq. (\ref{Zeld_Peebles_eq1}) of {\bf atlant} are 
calculated at the temperature of matter $T_{m}$] and {\bf recfast}. 
Dashed and dotted curves are partially overlapped.}
\label{fig1}
\end{figure}

Note that recombination and ionization coefficients included 
in the kinetic equation (\ref{Zeld_Peebles_eq1}) should be calculated at 
different temperatures (see e.g. \cite{Ali_Haimoud_Hirata10b}). 
Since the kinetics of recombination process is 
determined by the free electron distribution function, the recombination 
coefficient depends on kinetic temperature of free electrons $T_{m}$. 
The kinetics of ionization is determined by the photon distribution function, 
therefore ionization coefficient depends on temperature of photons $T$. 
Thus the detailed balance does not take place in considered case, and relation 
(\ref{hydrogen_ioniz_coeff}) has mathematical sense only and allows us 
to avoid direct calculation of ionization coefficient, 
$\beta_{HI}\left(T\right)$, from integral of collisions 
for photons and hydrogen atoms. Also exponential term in ionization part 
of (\ref{Zeld_Peebles_eq1}), 
$\exp\left(-{h\nu_{\alpha}/ \left[k_{B}T\right]}\right)$, should be calculated 
at the temperature of radiation. 
The distinction of temperatures used for calculation of ionization 
coefficients in the present code and in {\bf recfast} (\cite{Seager_et_al99}; 
\cite{Wong_et_al08}, there the temperature of matter is used for this aim) 
leads to a little but important difference in the free electron fraction 
(see Fig. \ref{fig1}). 

The equation (\ref{Zeld_Peebles_eq1}) is solved numerically together with 
other main kinetic equations (\ref{kin_eq_HeII}), (\ref{kin_eq_for_HeIII}) 
by using second-order method of integration of ODE system. The relative 
deviation between results obtained by {\bf atlant} and {\bf recfast} 
for the period of hydrogen cosmological recombination is presented in 
Fig. \ref{fig1}.


\subsection{Radiative Feedbacks}
\begin{figure}
\centering
\includegraphics[bb = 20 50 525 525, width=8cm, height=8cm]{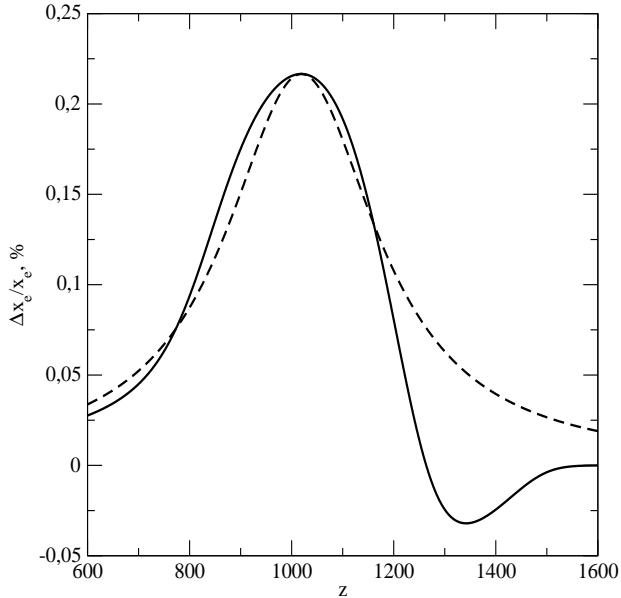}
\caption{Dashed line corresponds to the relative difference between 
{\bf atlant} with fudge function $\delta_{X}$ and base result 
[i.e. $\Delta x_{e}/x_{e}=\delta_{X}$]. Parameters of fudge function 
are the following: $A_{p}=2.166\cdot 10^{-3}$, 
$z_{p}=1019$, $\Delta z_{p}=180$. 
For comparison: 
solid line corresponds to the relative difference between feedback result 
for $n_{\rm max}=10$ [i.e. feedbacks 
$11\Rightarrow ... \Rightarrow 2$ are taken into account] 
and the base result [i.e. without feedbacks].}
\label{fig3}
\end{figure}

Due to a great number of fine effects the radiative feedbacks for resonant 
transitions have been chosen for including in the first published version of 
recombination code. It is because physics of this effect is clear 
(it is difficult to state this is true about many other fine effects) and 
independently obtained results of calculation of this effect 
(\cite{Chluba_Sunyaev10a, Kholupenko_et_al10}) confirm each other. 
Inclusion of feedbacks into the code is based on formulas suggested 
by \cite{Kholupenko_et_al10}. According this the total effective coefficient 
of np$\leftrightarrow$1s transitions is: 
\begin{equation}
A^{r}_{eff}={8\pi H \nu_{\alpha}^3 \over N_{HI} c^3}\delta_{A}
\label{A_r_eff}
\end{equation}
where $\delta_{A}$ is given by the following expression
\begin{equation}
\delta_{A}=\sum_{n\ge 2}^{n_{max}}C_{n}{\nu_{n}^3 \over \nu_{\alpha}^3}
\exp\left(-{E_{n}-E_{2}\over k_{B}T}\right)
\label{delta_A_r}
\end{equation}
where $\nu_{n}$ is the frequency of $n\rightarrow 1$ transition
($\nu_{2}=\nu_{\alpha}$), $E_n=h\nu_{n}$, and coefficients $C_{n}$ are:
\begin{equation}
C_{n}=\left(1-\Gamma_{H}(z'_{n})/\Gamma_{H}(z)\right)
\label{C_n_main}
\end{equation}
where in turn $\Gamma_{H}$ is the relative overheating of Ly$\alpha$ 
radiation (occupation number $\eta_{\alpha}$) in comparison with its 
equilibrium value (occupation number $\eta^{0}_{\alpha}$):
\begin{equation}
\Gamma_{H}=\left({\eta_{\alpha} / \eta^{0}_{\alpha}} - 1\right)
\label{Gamma_H}
\end{equation}
and
\begin{equation} 
z'_{n}=\left((1+z){\nu_{n+1}/\nu_{n}}-1\right)
\label{z_n}
\end{equation}
The set of formulas (\ref{A_r_eff}-\ref{z_n}) allows us to calculate 
ionization history taking into account the feedbacks for hydrogen 
$n\rightarrow 1$ resonant transitions with principal quantum numbers 
$n\le n_{max}$. For taking into account the feedback effect we use 
simple perturbation theory: at the first stage (unperturbed equations) 
{\bf atlant} calculates the relative overheating $\Gamma_{H}$ and stores this, 
at the second stage {\bf atlant} solves equations perturbed by the feedback 
effect. Relative deviation of free electron fraction between perturbed 
($n_{max}=10$) and unperturbed calculations is presented in Fig. \ref{fig3}.

\subsection{Fudge factors}
Developed recombination code allows users to use fudge factors as well as 
add physical effects. This opportunity has been included because of the 
following reasons: 1) list of fine effects leading to 0.01 - 0.1 \% corrections 
in the ionization fraction may be incomplete, and it is difficult to say 
how long this list will be in final form and when this will be achieved; 
2) considerations of some fine effects give contradictory results due 
to not completely clear physical picture (e.g. two-photon transitions from 
high excited states [\cite{Wong_Scott07}, \cite{Hirata08}, and 
\cite{Labzowsky_et_al09}], recoil [\cite{Grachev_Dubrovich08} and 
\cite{Hirata_Forbes09}] and others).

First fudge factor is the common (see {\bf recfast}) hydrogen fudge factor 
$F_{H}$ introduced in the expression (\ref{hydrogen_recomb_coeff}) for 
hydrogen recombination coefficient $\alpha_{HII}$.

Second fudge factor is the perturbation function $\delta_{X}$ modifying 
free electron fraction: 
\begin{equation}
x_{e}^{res}=x_{e}^{calc}\left(1+\delta_{X}\right)
\end{equation}
where $x_{e}^{res}$ is the free electron fraction (normalized by the 
total concentration of hydrogen atoms and ions) being the final result 
of code running (i.e. it is value shown in the resulting file), 
$x_{e}^{calc}$ is the solution of ODE's system 
describing ionization fractions and free electron fraction. 

Note that function $\delta_{X}$ affects final result $x_{e}$ but not ODE's 
system. This allows user to control changes of free electron fraction strictly 
by including fudge factors. 

Analyzing previous works devoted to the fine effects of cosmological 
recombination (e.g. \cite{Grachev_Dubrovich08, Chluba_Sunyaev09a}) 
one may note that typical form of corrections 
to the ionization fraction can be described by a bell-shaped function 
(e.g. Lorentzian or Gaussian profile or others). In this work the 
Lorentzian function has been chosen to describe uncertain deviations 
(until now) of free electron fraction from well known ODE's solution:
\begin{equation}
\delta_{X}={A_{p} \over 1+\left[\left(z-z_{p}\right)/\Delta z_{p}\right]^2}
\end{equation}
where $A_{p}$ is the relative amplitude of perturbation of ionization fraction, 
$z_{p}$ is the redshift of perturbation maximum, $\Delta z_{p}$ is the 
half-width of perturbation function at half-altitude.

Result of use of fudge function is shown in the Fig. \ref{fig3} where 
we have plotted $\Delta x_{e}/x_{e}=\delta_{X}$ as function of redshift $z$. 
Here we show how fudge function can mimic real fine corrections by means 
of example of feedback correction for $n_{max}=10$.

\section{HeII$\rightarrow$HeI helium recombination}
The time-dependent behaviour of HeII fraction in the isotropic 
homogeneous expanding Universe is described by the following kinetic 
equation (\cite{Kholupenko_et_al07, Wong_et_al08}): 
\begin{eqnarray}
\dot x_{HeII}=-C_{par}
[\alpha_{par}N_e x_{HeII} - {g_{a} \over g_{g}}\beta_{par}
\exp{\left(-{E_{ag}\over k_BT}\right)}x_{HeI} ] \nonumber  \\ 
-C_{or} [\alpha_{or}N_e x_{HeII} - {g_{a'} \over g_{g} }\beta_{or}
\exp{\left(-{E_{a' g}\over k_BT}\right)}x_{HeI} ]
\label{kin_eq_HeII}
\end{eqnarray} 
where $x_{HeII}=N_{HeII}/\left(N_{H}+N_{He}\right)$ is the fraction of 
HeII ions relative to the total number of hydrogen and helium atoms and ions, 
$C_{par}$ is the factor by which the ordinary 
recombination rate is inhibited by the presence of HeI $2^1p\rightarrow 1^1s$ 
resonance-line radiation, $\alpha_{par}$ is the total HeII$\rightarrow$HeI 
recombination coefficient to the excited para-states of HeI, subscript $a$ 
denotes $2^1s$ state of HeI atom, $N_e$ is the free electron concentration, 
$g_{a}=1$ is the statistical weight of $2^1s$ state of HeI, $g_{g}=1$ is the 
statistical weight of $1^1s$ state of HeI, $\beta_{par}$ is the total 
HeI$\rightarrow$HeII ionization coefficient from the excited para-states of 
HeI, $E_{ag}$ is the $2^1s\rightarrow 1^1s$ transition energy, 
$C_{or}$ is the factor by which the ordinary 
recombination rate is inhibited by the presence of HeI $2^3p\rightarrow 1^1s$ 
resonance-line radiation, $\alpha_{or}$ is the total HeII$\rightarrow$HeI 
recombination coefficient to the excited ortho-states of HeI, 
subscript $a'$ denotes state $2^3s$ of HeI atom $g_{a'}=3$ is the 
statistical weight of $2^3s$ state of HeI, $\beta_{or}$ is the total 
HeI$\rightarrow$HeII ionization coefficient from the excited ortho-states of 
HeI, $E_{a'g}$ is the $2^3s\rightarrow 1^1s$ transition energy, 
$x_{HeI}=N_{HeI}/\left(N_{H}+N_{He}\right)$ is the neutral helium fraction.

The para- and ortho- recombination coefficients are given by widely used 
approximation formulas (e.g. \cite{Verner_Ferland96}) parameters of which 
are based on data by \cite{Hummer_Storey98}):
\begin{equation}
\alpha_{S}\left(T\right)=q_{S}\left({T\over T_{2}}\right)^{-1/2}
\left(1+{T\over T_{2}}\right)^{-(1-p_{S})}
\left(1+{T\over T_{1}}\right)^{-(1+p_{S})}
\end{equation}
where subscript $S$ takes values `par' or `or', and $q_{S}$, $p_{S}$, 
$T_{1}=10^{5.114}$ K, $T_{2}=3$ K (\cite{Seager_et_al99}) 
are the parameters of approximation. For the recombination via para-states 
we have $q_{par}=10^{-10.744}$cm$^{3}$s$^{-1}$, $p_{par}=0.711$ 
(\cite{Seager_et_al99}). For the recombination via ortho-states we have 
$q_{or}=10^{-10.306}$cm$^{3}$s$^{-1}$, $p_{or}=0.761$ (\cite{Wong_et_al08}).

The para- and ortho- recombination and ionization coefficients are related 
by the following formula
\begin{equation}
\beta_{S}={g_c \over g_{S}}\alpha_{S} g_e(T) 
\exp\left({-{E_{S} \over k_B T}}\right), 
\label{HeI_detailed_balance}
\end{equation}
where subscript $``c''$ denotes continuum state of HeI atom, 
$g_c=4$ is the statistical weight of continuum state of (He$^+$+e$^-$), 
$E_{par}$ is the $c \rightarrow 2^1s$ transition energy, 
$E_{or}$ is the $c \rightarrow 2^3s$ transition energy. 

The inhibition factor $C_{par}$ is given by the following expression:
\begin{equation}
C_{par}={\left(g_{b}/g_{a}\right)A_{bg}P_{bg}
\exp\left(-{E_{ba} / k_BT}\right)+A_{ag}
\over \beta_{par}+\left(g_{b}/g_{a}\right)A_{bg}P_{bg}
\exp\left(-{E_{ba} / k_BT}\right)+A_{ag}}
\end{equation}
where $A_{bg}$ is the Einstein coefficient [s$^{-1}$] of 
$2^1p\leftrightarrow 1^1s$ spontaneous transitions, 
$P_{bg}$ is the probability of the uncompensated $2^1p\rightarrow 1^1s$
transitions,
$E_{ba}$ is the $2^1p\rightarrow 2^1s$ transition energy, 
$A_{ag}$ is the coefficient of two-photon $2^1s\rightarrow 1^1s$ spontaneous 
decay.

The inhibition factor $C_{or}$ is given by the following expression 
\begin{equation}
C_{or}={\left(g_{b'} / g_{a'} \right)A_{b'g}P_{b'g}
\exp\left(-{E_{b'a'}/ k_BT}\right) 
\over \beta_{or}+\left(g_{b'} / g_{a'} \right)A_{b'g}P_{b'g}
\exp\left(-{E_{b'a'}/ k_BT}\right)}
\end{equation}
where subscript $b'$ denotes state $2^3p$ of HeI, 
$g_{b'}=9$ is the statistical weight of $2^3p$ state of HeI, 
where $A_{bg}$ is the Einstein coefficient [s$^{-1}$] of 
$2^3p\leftrightarrow 1^1s$ spontaneous transitions, 
$P_{b'g}$ is the probability of the uncompensated $2^3p\rightarrow 1^1s$
transitions,
$E_{b'a'}$ is the $2^3p\rightarrow 2^3s$ transition energy. 

The probabilities $P_{bg}$ and $P_{b'g}$ take into account the escape of 
HeI resonant photons from the line profiles due to the cosmological 
expansion and destruction of these photons by neutral hydrogen. 
They can be found by the following formula:
\begin{equation}
P_{fg}=P^{r}_{fg}+P^{H}_{fg}
\end{equation}
where $P^{r}_{fg}$ is approximately given by the following: 
\begin{equation}
P^{r}_{fg}=\left(1+\gamma^{-1}\right)^{-2}
\tau_{He,f}^{-1}\left(1-\exp\left(-\tau_{He,f}\right)\right)
\label{modified_Sobolev_probability}
\end{equation}
where $\gamma$ is the ratio of the helium and
hydrogen absorption coefficients at the central frequency
of the $f\rightarrow g$ line (here symbol $f=b$ or $b'$ depending on what 
transition is considered), $\tau_{He,f}$ is the Sobolev optical depth. 
The value $\gamma$ is given by the following relation: 
\begin{equation}
\gamma = {\left(g_{f}/g_{g}\right) A_{fg} N_{HeI} c^2 \over 
\sigma_{H}\left(\nu_{fg}\right) 8\pi^{3/2}
\nu_{fg}^2 \Delta \nu_{D,f} N_{HI}}
\label{gamma_f}
\end{equation}
where $\sigma_{H}$ is the ionization cross-section of hydrogen ground state, 
parameter $\Delta \nu_{D,f}=\nu_{fg}\sqrt{2k_{B}T/\left(m_{He}c^2\right)}$ 
is the Doppler line width.

\begin{table}
\centering
\caption{Parameters of the approximation of $P^{H}_{D}$}
\begin{tabular}{lcc}
  \hline
  Range of $\gamma$ & p & q \\
  \hline
  $0\le \gamma \le 5\cdot 10^{2}$ & 0.66 & 0.9 \\
  $5\cdot 10^{2} < \gamma \le 5\cdot 10^{4}$ & 0.515 & 0.94 \\
  $5\cdot 10^{4} < \gamma \le 5\cdot 10^{5}$ & 0.416 & 0.96 \\
  $5\cdot 10^{5} < \gamma $ & 0.36 & 0.97 \\
  \hline
  \label{tab_pq}
\end{tabular}
\end{table}

The optical depth $\tau_{He,f}$ is:
\begin{equation}
\tau_{He,f}={g_{f}A_{fg}N_{HeI}c^3/\left(g_{g}8\pi H\nu_{fg}^3\right)}
\label{Sobolev_opt_depth_He}
\end{equation}

The value $P^{H}_{fg}$ is given by the following:
\begin{equation}
P^{H}_{fg}=P^{H}_{D}+P^{H}_{G}+P^{H}_{R}
\end{equation}
where in turn $P^{H}_{D}$ approximately is:
\begin{equation}
P^{H}_{D}=\left(1+p\gamma^{q}\right)^{-1}
\label{P_H_D_approx}
\end{equation}
where parameters $p$, $q$ are given in the Tab. \ref{tab_pq}. 

The value $P^{H}_{G}$ is given by the following approximate formula:
\begin{eqnarray}
P^{H}_{G}=(8\lambda a)^{1/4}\pi^{-5/8}\gamma^{-3/4}\cdot \nonumber \\
\cdot\left[1+\exp\left(-1.07\ln\left(s+1.5\right)-0.45\right)\right] 
\left(s+1.28\right)^{-1/2} 
\label{P_H_G}
\end{eqnarray}
where $a=\Gamma_{f} / 4\pi \Delta \nu_{D,f}$ is the Voigt parameter 
(here $\Gamma_{f}$ is the natural line width), 
$\lambda$ is the single-scattering albedo in $f\rightarrow g$ line, 
and parameter $s$ is given by the following relation (\cite{Grachev88}): 
\begin{equation}
s=2^{-3/2}\pi^{-1/4}(1-\lambda)\lambda^{-1/2}
a^{1/2}\gamma^{1/2}-1/4
\end{equation}

The single-scattering albedo $\lambda$ can be calculated by:
\begin{equation}
\lambda\simeq {A_{fg}\over A_{fg}+R_{f}}
\end{equation}
where $R_{f}=\sum_{n}R_{f\rightarrow n}$ is the total coefficient of 
radiative transitions (including induced transitions, 
while collision transitions are considered negligible) 
from $f$ state to excited states of HeI atom.

\begin{figure}
\centering
\includegraphics[bb = 20 50 525 525, width=8cm, height=8cm]{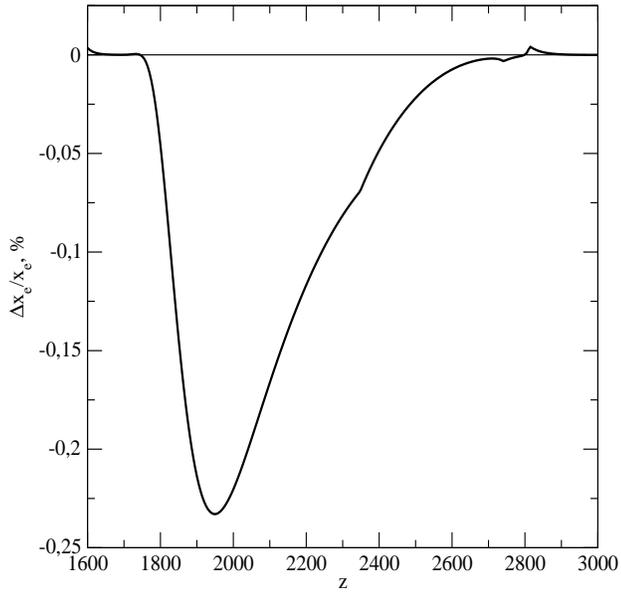}
\caption{The relative difference $\Delta x_{e}/x_{e}$ between free 
electron fractions calculated by {\bf atlant} and {\bf recfast} for 
the period of HeII$\rightarrow$HeI recombination.}
\label{fig4}
\end{figure}

The value $P^{H}_{R}$ is approximately:
\begin{equation}
P^{H}_{R}=2(1-\lambda)\sqrt{a \over (1-\lambda)\gamma \pi^{3/2}} 
\left({\pi \over 2} - \arctan \left(f(a, \lambda, \gamma)\right)\right)
\label{P_H_R2}
\end{equation}
where $f(a, \lambda, \gamma)$ is defined by the expression
\begin{equation}
f(a, \lambda, \gamma)\simeq 0.69\cdot
(1-\lambda)^{-1/2}\lambda^{1/4}\left( a \gamma\right)^{-1/4}\sqrt{\ln{\gamma}}
\label{f_limit_final}
\end{equation}
The equation (\ref{kin_eq_HeII}) is solved numerically together with 
other main kinetic equations (\ref{Zeld_Peebles_eq1}) and 
(\ref{kin_eq_for_HeIII}) to determine free electron fraction $x_{e}$. 
The relative deviation of free electron fraction between the results calculated 
by using {\bf atlant} and {\bf recfast} for the period of HeII$\rightarrow$HeI 
recombination is presented in Fig. \ref{fig4}.

\section{HeIII$\rightarrow$HeII helium recombination}
In the difference with {\bf recfast} we use non-equilibrium (i.e. kinetic) 
approach for consideration of HeIII$\rightarrow$HeII helium recombination. 
The time-dependent behaviour HeIII fraction in the isotropic 
homogeneous expanding Universe is described by the following kinetic 
equation:
\begin{eqnarray}
\dot x_{HeIII}=-C_{HeII}[\alpha_{HeIII}N_e x_{HeIII} - \nonumber \\
\beta_{HeII}\exp\left(-{h\nu_{HeII,21}\over k_{B}T}\right)x_{HeII}]
\label{kin_eq_for_HeIII}
\end{eqnarray} 
where $x_{HeIII}=N_{HeIII}/\left(N_{H}+N_{He}\right)$, $C_{HeII}$ is the factor 
by which the ordinary recombination rate is inhibited by the presence of 
HeII $2p\rightarrow 1s$ resonance-line radiation, $\alpha_{HeIII}$ is the total 
HeIII$\rightarrow$HeII recombination coefficient to the excited states of HeII, 
$\beta_{HeII}$ is the total HeII$\rightarrow$HeIII ionization coefficient from 
the excited states of HeII, $\nu_{HeII,21}$ is the HeII $2p\rightarrow 1s$ 
transition frequency.

The inhibition coefficient is given by the expression:
\begin{equation}
C_{HeII}={A^{r}_{HeII,2p1s} + A_{HeII,2s1s} \over 
\beta_{HeII}+A^{r}_{HeII,2p1s} + A_{HeII,2s1s}}
\label{C_HI_res}
\end{equation}
where $A^{r}_{HeII,2p1s}$ is the effective coefficient of 
HeII 2p$\leftrightarrow$1s transitions, $A_{HeII,2s1s}$ is the coefficient of 
HeII 2s$\rightarrow$1s two-photon spontaneous transition. 

Since HeII is the hydrogenic ion the recombination coefficient 
$\alpha_{HeIII}$ can be found by using simple scaling relation (see e.g. 
\cite{Verner_Ferland96}): 
\begin{equation}
\alpha_{Z}\left(T\right)=Z\alpha_{1}\left(T/Z^{2}\right)
\label{alpha_scaling}
\end{equation} 
where $Z$ is the nuclear charge for the hydrogenic ions. 
In considered case Eq. (\ref{alpha_scaling}) yields 
$\alpha_{HeIII}\left(T\right)=2\alpha_{HII}\left(T/4\right)$, where 
$\alpha_{HII}$ is taken from (\ref{hydrogen_recomb_coeff}) without 
hydrogen fudge factor $F_{H}$. 

The ionization coefficient $\beta_{HeII}$ can be found by using principle of 
detailed balance:
\begin{equation}
\beta_{HeII}\left(T\right)=\alpha_{HeIII}\left(T\right)
g_e\left(T\right)\exp\left({-{h\nu_{HeII,c2} \over k_B T}}\right)
\label{HeII_ioniz_coeff}
\end{equation}
$\nu_{HeII,c2}$ is the HeII c$\rightarrow$2 transition frequency. 

Note that in Eq (\ref{kin_eq_for_HeIII}) the recombination and ionization 
coefficients are calculating at the same temperature $T$. This is because 
the matter temperature $T_{m}$ is very close to the radiation one $T$ 
(relative deviation is less than $10^{-5}$) during HeIII$\rightarrow$HeII 
recombination and calculation at different temperatures has no sense at 
required level of accuracy.

The effective coefficient $A^{r}_{HeII,2p1s}$ of HeII 2p$\leftrightarrow$1s 
transitions due to escape of HeII 2p$\leftrightarrow$1s resonant photons from 
the line profile because of cosmological redshift is given by the following 
formula:
\begin{equation}
A^{r}_{HeII,2p1s}={8\pi H \nu_{HeII,21}^3 \over N_{HeII} c^3}
\label{A_r_HeII}
\end{equation}

The value $A_{HeII,2s1s}$ is found from charge scaling for the 
hydrogenic ions $A_{Z,2s1s}=Z^6A_{2s1s}$ (see e.g. \cite{Shapiro_Breit59}, 
\cite{Zon_Rapoport68}, \cite{Nussbaumer_Schmutz84}). In the considered case 
this gives us $A_{HeII,2s1s}=2^6A_{2s1s}$. 

\begin{figure}
\centering
\includegraphics[bb = 20 50 525 525, width=8cm, height=8cm]{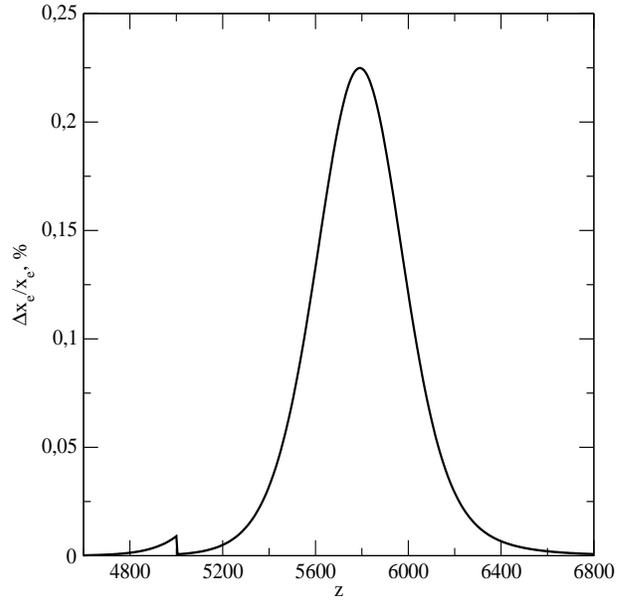}
\caption{The relative difference $\Delta x_{e}/x_{e}$ between free 
electron fractions calculated by {\bf atlant} and {\bf recfast} for the period of 
HeIII$\rightarrow$HeII recombination.}
\label{fig5}
\end{figure}

The equation (\ref{kin_eq_for_HeIII}) is solved numerically together with 
other main kinetic equations (\ref{Zeld_Peebles_eq1}) and 
(\ref{kin_eq_HeII}) to determine free electron fraction $x_{e}$. 
The relative deviation between results by {\bf atlant} and {\bf recfast} 
for the period of HeIII$\rightarrow$HeII recombination is presented 
in Fig. \ref{fig5}.

\section{Evolution of matter temperature}
The behaviour of the matter temperature in the isotropic 
homogeneous expanding Universe is described by 
the following equation (e.g. \cite{Peebles68, Scott_Moss09}):
\begin{equation}
\dot T_{m}={8\sigma_{T}a_{R}T^{4}\over 3m_{e}c}
{x_{e}\over 1+x_{e}+N_{He0}/N_{H0}}\left(T-T_{m}\right)-2HT_{m}
\label{matter_temp_eq}
\end{equation}
where $\sigma_{T}$ is the Thomson scattering cross section.

Defining relative deviation, $\delta_{T}$, of the matter temperature from 
the radiation temperature via $T_{m}=T\left(1-\delta_{T}\right)$ and 
substituting this into (\ref{matter_temp_eq}) one can obtain:
\begin{equation}
\dot \delta_{T} = -\left(R_{T} + H\right)\delta_{T} + H
\label{matter_temp_delta}
\end{equation}
where $R_{T}$ is the rate of energy transfer between matter and radiation 
via Compton scattering:
\begin{equation}
R_{T}={8\sigma_{T}a_{R}T^{4}\over 3m_{e}c}{x_{e}\over 1+x_{e}+N_{He0}/N_{H0}}
\end{equation}

To solve equation (\ref{matter_temp_delta}) we applied a perturbation 
approach. In the early stages of the Universe history ($z\gtrsim 200$) 
the rate of energy transfer between matter and radiation via Compton 
scattering is much larger than the rate of the temperature change (the latter 
is about Hubble expansion rate $H$), i.e. 
$\dot\delta_{T}/\left(R_{T}\delta_{T}\right)\ll 1$. 
Thus we can use expansion of the solution over this smallness: 
\begin{equation}
\delta_{T}=\sum_{i=0}^{\infty}\delta_{T,i}
\label{delta_T_series}
\end{equation}
where zeroth-order approximation is determined as quasistacionary solution 
of Eq. (\ref{matter_temp_delta}) 
(see also \cite{Hirata08, Ali_Haimoud_Hirata10a}): 
\begin{equation}
\delta_{T,0}=\left(R_{T}/H+1\right)^{-1}
\label{delta_T0}
\end{equation}
and the next members of expansion (\ref{delta_T_series}) are related by 
the following equation:
\begin{equation}
\delta_{T,i+1}=-\left(R_{T} + H\right)^{-1}\dot\delta_{T,i}
\label{delta_Tm}
\end{equation}
\begin{figure}
\centering
\includegraphics[bb = 20 50 525 525, width=8cm, height=8cm]{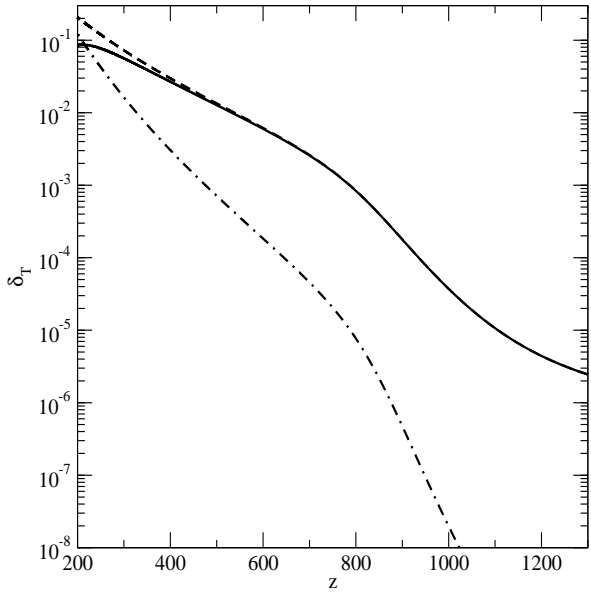}
\caption{The corrections, $\delta_{T}$, to the matter temperature, $T_{m}$,
relative to the radiation temperature, $T$, as functions of redshift $z$: 
dashed curve corresponds to the correction, $\delta_{T,0}$, dashed-dotted 
curve corresponds to  $|\delta_{T,1}|$ ($\delta_{T,1}<0$), 
and solid curve corresponds to 
$\left(\delta_{T,0}+\delta_{T,1}\right)$.}
\label{fig6}
\end{figure}
In the present version of the code we keep only two first corrections 
$\delta_{T,0}$ and $\delta_{T,1}$ in Eq. (\ref{delta_T_series}).

For the period $z\lesssim 200$ the quasistationary condition is violated, 
so expansion (\ref{delta_T_series}) loses convergence and cannot represent 
the solution of Eq. (\ref{matter_temp_delta}). In this redshift range we 
use the following dependence of matter temperature on redshift:
\begin{equation}
T_{m}=T_{m}^{dec}\left({1+z \over 1+z_{dec}}\right)^{2}
\label{usual_matter_temp}
\end{equation}
where $T_{m}^{dec}$ is the matter temperature from (\ref{delta_T_series}) at 
$z_{dec}$ (index ``dec'' means ``decoupling'') which is considered as 
the moment of decoupling of the matter temperature from radiation the one 
(at value $z_{dec}$ we should make join of solutions). In present version of code 
we determine $z_{dec}$ from condition $\delta_{T,0}=0.5$. Equation 
(\ref{usual_matter_temp}) corresponds to the equation of state 
of non-relativistic matter.

\begin{figure}
\centering
\includegraphics[bb = 20 50 525 525, width=8cm, height=8cm]{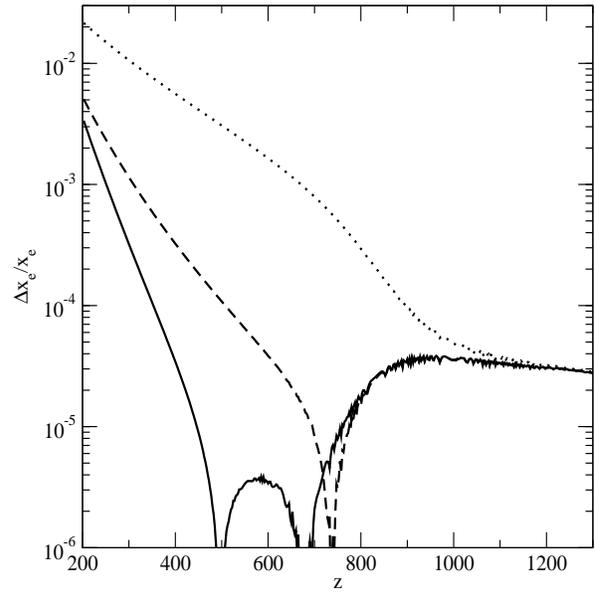}
\caption{The relative difference $\Delta x_{e}/x_{e}$ of the free 
electron fractions as a function of redshift $z$: dotted curve 
corresponds to the difference, $\Delta x_{e}/x_{e}$, between results for $\delta_{T}=0$ 
(i.e. $T_{m}=T$) by {\bf atlant} and {\bf recfast} corrected (for the details of 
corrections, see section \ref{Hydrogen_Main_equations} of the present paper),  
dashed curve corresponds to the difference, $\Delta x_{e}/x_{e}$, between results for 
$\delta_{T}=\delta_{T,0}$ by {\bf atlant} and corrected {\bf recfast}, 
and solid curve corresponds to the difference, $\Delta x_{e}/x_{e}$, between results for 
$\delta_{T}=\left(\delta_{T,0}+\delta_{T,1}\right)$  by {\bf atlant} 
and corrected {\bf recfast}.}
\label{fig7}
\end{figure}

Results of calculations of $\delta_{T,0}$ and $\delta_{T,1}$ are presented 
in Fig. \ref{fig6}. The influence of taking into account different 
approximations for $\delta_{T}$ [depending on the number of kept members of 
expansion (\ref{delta_T_series})] 
on the free electron fraction is shown in Fig. \ref{fig7}.

\section{Variation of the fundamental constants}
Today the opportunity to vary of the fundamental constants becomes essential 
at the analysis of CMBR anisotropy (e.g. \cite{Scoccola_et_al08}) and we 
decided to include this in our code. Thus the current version of the code 
allows user to see how recombination occurs at different values of the 
fundamental constants (this means that changes of fundamental constants 
lead to corresponding changes of derived physical 
values, e.g. ionization energies of atoms, Thomson cross section and others). 

Since some of used physical values are given only numerically 
(e.g. level energies and transition probabilities for HeI atom, 
see www.nist.gov) 
we use simple scalings to take into account the influence of variation 
of fundamental constants on these values. These scalings are the following: 
\\1) For the level energies:
\begin{equation}
E=E_{st}\left({e \over e_{st}}\right)^{4}
\left({m_{e} \over m_{e,st}}\right)
\left({\hbar \over \hbar_{st}}\right)^{-2}
\label{energy_scaling}
\end{equation}
2) For the one-photon transition coefficients:
\begin{equation}
A^{\gamma}=A^{\gamma}_{st}\left({e \over e_{st}}\right)^{10}
\left({m_{e} \over m_{e,st}}\right)
\left({\hbar \over \hbar_{st}}\right)^{-6}
\left({c \over c_{st}}\right)^{-3}
\label{one_photon_trans_scaling}
\end{equation}
3) For the two-photon transition coefficients:
\begin{equation}
A^{\gamma\gamma}=A^{\gamma\gamma}_{st}\left({e \over e_{st}}\right)^{16}
\left({m_{e} \over m_{e,st}}\right)
\left({\hbar \over \hbar_{st}}\right)^{-9}
\left({c \over c_{st}}\right)^{-6}
\label{two_photon_trans_scaling}
\end{equation}
4) For the recombination coefficients:
\begin{equation}
\alpha=\alpha_{st}\left({e \over e_{st}}\right)^{6}
\left({m_{e} \over m_{e,st}}\right)^{-3/2}
\left({\hbar \over \hbar_{st}}\right)^{-1}
\left({c \over c_{st}}\right)^{-3}
\label{recombination_scaling}
\end{equation}
In the expressions (\ref{energy_scaling})-(\ref{recombination_scaling}) 
subscript ``st'' denotes current values of physical quantities while 
symbols without subscript denote values which user suggests to be valid 
during the recombination epoch.

\begin{figure}
\centering
\includegraphics[bb = 20 50 525 525, width=8cm, height=8cm]{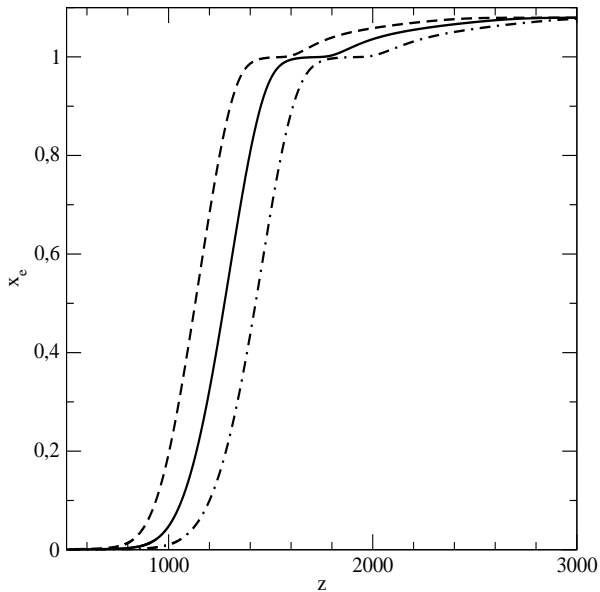}
\caption{Free electron fractions $x_{e}$ as functions of redshift $z$ 
at different values of the fine-structure constant $\alpha_{fine}$: 
dashed curve corresponds to $\alpha_{fine}/\alpha^{0}_{fine}=0.95$, 
solid curve corresponds to $\alpha_{fine}/\alpha^{0}_{fine}=1$,
dashed-dotted curve corresponds to $\alpha_{fine}/\alpha^{0}_{fine}=1.05$}
\label{fig8}
\end{figure}

The influence of variations of the fine structure constant $\alpha_{fine}$ 
(via varying elementary charge) on the free electron fraction is shown 
in Fig. \ref{fig8}.

\section{Results}
The main numerical results of this paper are the relative differences 
of free electron fractions $\Delta x_{e} /x_{e}$ calculated by {\bf atlant} 
and {\bf recfast} (figs \ref{fig1}, \ref{fig4}, \ref{fig5}). 
In the top panel of Fig. \ref{fig1} the relative difference 
$\Delta x_{e} /x_{e}$ between result of {\bf atlant} and current 
version of {\bf recfast} (version 1.5, \cite{Wong_et_al08}) for the 
hydrogen recombination epoch. 
The first significant difference appears in the period $z = 1550 - 1650$. 
This is because at early epochs $z\gtrsim 1550$ {\bf recfast} considers 
recombination as quasi-equilibrium process (according to Saha formula) and 
sets initial condition for the kinetic equation correspondingly, while 
{\bf atlant} considers recombination as a non-equilibrium (kinetic) 
process in the whole range of redshifts and sets an initial conditions 
corresponding to the fully ionized plasma at the moment given by user 
(in our case $z_{begin}=8000$). 
The sharp peak of $\Delta x_{e} /x_{e}$ in the period $z = 1550 - 1650$ 
(maximal value is about $5\cdot 10^{-4}$)
is due to transition from recombination according to Saha formula 
to non-equilibrium recombination which occurs in {\bf recfast} at 
free electron fraction $x_{e}=0.985$ (that corresponds to $z\simeq 1555$ in 
considered case). There is a simple method of modelling of CMBR spectral 
distortion due to cosmological recombination (e.g. \cite{Bernshtein_et_al77, 
Dubrovich_Stolyarov95, Dubrovich_Stolyarov97, Burgin03}). 
This method is based on the formalism of matrix of efficiency 
of radiative transitions (ERT-matrix) and it does not demand direct 
calculation of atomic level populations. On the other hand this method 
demands use of the following values $dx_{HII}/dz$, 
$dx_{HeII}/dz$, and $dx_{HeIII}/dz$ to determine the rates of resonance 
photon emission and correspondingly the shape of cosmological recombination 
lines. Thus for using ERT-matrix method the sharp numerical 
features may be significant defects of cosmological recombination modelling. 
The second difference of results by {\bf atlant} and {\bf recfast} appears 
in the period $z = 300 - 1000$. It arises due to different structure of 
ionization items of hydrogen kinetic equation (\ref{Zeld_Peebles_eq1}) in 
{\bf atlant} and {\bf recfast}: in {\bf recfast} the ionization coefficient 
$\beta_{HI}$ and Boltzmann exponential term are calculated at the temperature 
of matter while in {\bf atlant} at the radiation temperature. Most part of this 
difference is due to distinction of ionization coefficients including 
in the denominator of inhibition coefficient $C_{HI}$. The maximum of 
$\Delta x_{e} /x_{e}$ in mentioned range of redshifts is about 0.27\% 
at $z\simeq 770$. This is little but maybe important difference in the 
context of future analysis of Planck data. The third difference appears 
at low redshifts ($z\lesssim 300$). It is due to the different approaches 
to evaluating matter temperature in {\bf recfast} (where ODE for temperature 
is solved) and in {\bf atlant} (where perturbation theory for temperature 
estimate is used). In the bottom panel of Fig. \ref{fig1} 
we present three curves. Solid curve 
is the same as in the top panel but in a logarithmic scale. Dashed curve 
corresponds to the relative difference $\Delta x_{e}/x_{e}$ between 
free electron fractions calculated by {\bf atlant} and {\bf recfast} 
with corrected ionization rate [i.e. ionization coefficient 
$\beta_{HI}\left(T\right)$ and exponential term 
$\exp\left(-{h\nu_{\alpha}/\left[k_{B}T\right]}\right)$ in Eq. 
(\ref{Zeld_Peebles_eq1}) of the {\bf recfast} model are calculated at the 
temperature of radiation, $T$]. Dotted curve corresponds to the relative 
difference $\Delta x_{e}/x_{e}$ between free electron fractions calculated 
by {\bf atlant} with modified ionization rate [i.e. ionization coefficient 
$\beta_{HI}\left(T_{m}\right)$ and exponential term 
$\exp\left(-{h\nu_{\alpha}/\left[k_{B}T_{m}\right]}\right)$ in Eq. 
(\ref{Zeld_Peebles_eq1}) of the {\bf atlant} model are calculated 
at the temperature of matter $T_{m}$] and {\bf recfast}. These curves 
show that in the frame of identical physical models the accordance of 
results by {\bf atlant} and {\bf recfast} is wholly satisfactory. 
Residual difference which does not exceed $5\cdot 10^{-5}$ for the 
period $z = 400 - 1500$ can be explained by a little different values of 
physical constants and different integration methods which have been used 
in {\bf atlant} and {\bf recfast}. 

In Fig. \ref{fig3} we demonstrate how feedbacks for resonant transitions 
of hydrogen affect free electron fraction (solid curve) and show an example of 
using fudge function, $\delta_{X}$ (dashed curve). The change due to 
feedbacks has the maximum about 0.2166\% at redshift $z\simeq 1019$ in full 
accordance with \cite{Chluba_Sunyaev10a}. 
This calculation has been done for $n_{max}=10$. 
Parameters of the fudge function have been choosen as following: 
$A_{p}=2.166\cdot 10^{-3}$, $z_{p}=1019$, and $\Delta z_{p}=180$. This is 
to show how use of the fudge function allows us to imitate the influence of 
real physical effects on the free electron fraction.

In Fig. \ref{fig4} the relative difference of the results by {\bf atlant} 
and {\bf recfast} for the epoch of HeII$\rightarrow$HeI recombination is 
presented. The breaks in the range $z = 2700 - 2900$ and break at 
$z\simeq 2300$ are the numerical features connected with switch in 
{\bf recfast} code. The maximum of difference is about -0.23\% 
at $z\simeq 1950$. The difference between results by {\bf atlant} and 
{\bf recfast} for the epoch of HeII$\rightarrow$HeI recombination arises 
because different approximations for estimate 
of effective eacape probability are used: in {\bf atlant} the approach 
based on analytical consideration of resonance radiation transfer 
in the presence of 
continuum absorption is used (\cite{Kholupenko_et_al08}) while {\bf recfast} 
uses the simple approximation with an additional fudge factor $b_{He}$ 
(\cite{Wong_et_al08}). It leads to the numerical 
differences because used approximations are not identical. 
Also there is physical reason for divergence of these 
approximations: \cite{Wong_et_al08} have determined $b_{He}$ 
from the best agreement with result by \cite{Switzer_Hirata08a} who took 
into account not only hydrogen continuum opacity but also feedbacks for 
resonant transitions of HeI and found this effect is about 0.46\% at 
$z\simeq 2045$. Later works (e.g. \cite{Chluba_Sunyaev10a}) show that 
this feedback effect is about 0.17\% at $z\simeq 2300$, i.e. negligible for 
the most of modern problems connected with cosmological recombination 
(e.g. CMBR anisotropy analysis). Thus we do not take helium feedbacks 
into account in the present version of the code. 
One more important point of calculations 
of effective escape probability for HeI is the estimate of number of neutral 
hydrogen at the high redshifts $z\gtrsim 1600$ (i.e. when hydrogen ionization 
fraction close to unity [e.g. greater than 0.985]): {\bf recfast} uses Saha 
formula for this aim, while in {\bf atlant} the kinetic equation is solved 
in the whole range of redshifts to determine the fractions of all considered 
plasma components. This explains the discrepancy between results of the present 
paper and \cite{Kholupenko_et_al08} where approximation by \cite{Wong_et_al08} 
has also been investigated by using earlier version of our code 
(i.e. Wong-Moss-Scott approximation has been calculated, but 
number of neutral hydrogen has been calculated from kinetic equation). 

In Fig. \ref{fig5} the relative difference of the results by {\bf atlant} 
and {\bf recfast} for the epoch of HeIII$\rightarrow$HeII recombination is 
presented. The break at $z=5000$ is the numerical feature 
connected with switches in {\bf recfast} code again. 
In the range of redshifts $z = 5000 - 7000$ the difference 
between the results by {\bf atlant} and {\bf recfast} arises because 
HeIII$\rightarrow$HeII recombination is treated by {\bf atlant} in 
non-equilibrium way while {\bf recfast} treats this according to Saha formula. 
The maximum of difference is about 0.225\% at $z\simeq 5790$. Such difference 
is too small to be a reason of any observational effects at the current 
level of experimental accuracy, but non-equilibrium treatment of 
HeIII$\rightarrow$HeII allows us to avoid artificial numerical features 
(that is important for the modelling of CMBR spectral distortion arising from 
HeIII$\rightarrow$HeII recombination). 

In Fig. \ref{fig6} the deviation of the matter temperature from the 
radiation temperature is presented. The dashed curve corresponds to 
the correction $\delta_{T,0}$, dashed-dotted curve corresponds to $|\delta_{T,1}|$, 
and solid curve corresponds to $\left(\delta_{T,0}+\delta_{T,1}\right)$. 
One can see that difference between the radiation and matter temperature is 
less than $10^{-4}$ down to $z\simeq 940$ and less than $10^{-3}$ down to 
$z\simeq 790$. Taking into account that $T_{m}$ is included in kinetic equation 
(\ref{Zeld_Peebles_eq1}) only via recombination coefficient 
$\alpha_{HII}\left(T_{m}\right)$ one can expect the similar changes 
in hydrogen ionization fraction at transition from approximation $T_{m}=T$ 
to more exact estimate of $T_{m}$. From accuracy point of view the correction 
$\delta_{T,1}$ becomes important beginning from $z\simeq 480$ where/when it achieves 
values about $10^{-3}$. At the low $z$ ($z\lesssim 300$) the correction 
$\delta_{T,1}$ is larger than 20\% of $\delta_{T,0}$ and should be taken into 
account for correct determination of residual ionization fraction as possible. 
In Fig. \ref{fig7} we show how including of $\delta_{T,i}$'s affects the 
free electron fraction. We have compared our results with the results by 
{\bf recfast} corrected in the following way: ionization items of Eq. 
(\ref{Zeld_Peebles_eq1}), i.e. ionization coefficient $\beta_{HI}$ and 
exponential term $\exp\left(-{h\nu_{\alpha}/\left[k_{B}T_{m}\right]}\right)$ 
are calculated at temperature of radiation $T$. The dotted curve corresponds 
to the relative difference $\Delta x_{e}/x_{e}$ of the result by {\bf atlant} 
for approximation $T_{m}=T$ (i.e. $\delta_{T}=0$) and the mentioned result by 
{\bf recfast}, the dashed one corresponds to $\Delta x_{e}/x_{e}$ at 
$\delta_{T}=\delta_{T,0}$, and the solid one does $\Delta x_{e}/x_{e}$ at 
$\delta_{T}=\left(\delta_{T,0}+\delta_{T,1}\right)$. From Fig. \ref{fig7} 
one can see that rough approximation $T_{m}=T$ is quite valid (at required 
level of accuracy) down to $z\simeq 670$ where $\Delta x_{e}/x_{e}$ achieves 
values about $10^{-3}$. Thus using this rough approximation biases the results 
not so dramatically as it seems when this approximation is used in 
{\bf recfast}. More advanced approximations are valid 
($\Delta x_{e}/x_{e}\le 10^{-3}$) down to $z\simeq 310$ 
($\delta_{T}=\delta_{T,0}$) and $z\simeq 250$ 
($\delta_{T}=\left(\delta_{T,0}+\delta_{T,1}\right)$).

Fig. \ref{fig8} illustrates one of the additional opportunities of 
{\bf atlant}: here the changes of free electron fraction at variation 
of the fine-structure constant are presented. Presented results are obtained 
for the following values of the fine-structure constant $\alpha_{fine}$: 
$0.95\alpha^{0}_{fine}$, $\alpha^{0}_{fine}$, and $1.05\alpha^{0}_{fine}$ 
(where $\alpha^{0}_{fine}$ is the current value of the fine-structure constant). 
Obtained results are very similar to the results by \cite{Scoccola_et_al08}. 

\vspace{0.3cm}
\hspace{-0.7cm}
{\bf Acknowledgements}
Authors are grateful to participants of Workshops ``Physics of 
Cosmological Recombination'' (Max Planck Institute for Astrophysics, 
Garching 2008 and University Paris-Sud XI, Orsay 2009) for 
useful discussions on cosmological recombination. 

This work has been partially supported by Ministry of Education and 
Science of Russian Federation (contract \# 11.G34.31.0001 with SPbSPU 
and leading scientist G.G. Pavlov), RFBR grant 11-02-01018a, and 
grant ``Leading Scientific Schools of Russia'' NSh-3769.2010.2. 
Balashev S.~A. also thanks the Dynasty Foundation.

\label{lastpage}

\begin{thebibliography}{}

\bibitem[\protect\citeauthoryear{{Ali-Ha{\"i}moud}, {Grin} \&
  {Hirata}}{{Ali-Ha{\"i}moud} et~al.}{2010}]{Ali_Haimoud_et_al10}
{Ali-Ha{\"i}moud} Y.,  {Grin} D.,    {Hirata} C.~M.,  2010, Phys. Rev. D, 82,
  123502

\bibitem[\protect\citeauthoryear{{Ali-Ha{\"i}moud} \&
  {Hirata}}{{Ali-Ha{\"i}moud} \& {Hirata}}{2010a}]{Ali_Haimoud_Hirata10b}
{Ali-Ha{\"i}moud} Y.,  {Hirata} C.~M.,  2010a, arxiv:1011.3758

\bibitem[\protect\citeauthoryear{{Ali-Ha{\"i}moud} \&
  {Hirata}}{{Ali-Ha{\"i}moud} \& {Hirata}}{2010b}]{Ali_Haimoud_Hirata10a}
{Ali-Ha{\"i}moud} Y.,  {Hirata} C.~M.,  2010b, Phys. Rev. D, 82, 063521

\bibitem[\protect\citeauthoryear{{Bernshtein}, {Bernshtein} \&
  {Dubrovich}}{{Bernshtein} et~al.}{1977}]{Bernshtein_et_al77}
{Bernshtein} I.~N.,  {Bernshtein} D.~N.,    {Dubrovich} V.~K.,  1977, Soviet
  Astronomy, 21, 409

\bibitem[\protect\citeauthoryear{{Boschan} \& {Biltzinger}}{{Boschan} \&
  {Biltzinger}}{1998}]{Boschan_Biltzinger98}
{Boschan} P.,  {Biltzinger} P.,  1998, A\&A, 336, 1

\bibitem[\protect\citeauthoryear{{Burgin}}{{Burgin}}{2003}]{Burgin03}
{Burgin} M.~S.,  2003, Astronomy Reports, 47, 709

\bibitem[\protect\citeauthoryear{{Chluba}, {Rubi{\~n}o-Mart{\'{\i}}n} \&
  {Sunyaev}}{{Chluba} et~al.}{2007}]{Chluba_et_al07}
{Chluba} J., {Rubi{\~n}o-Mart{\'{\i}}n} J.~A., {Sunyaev} R.~A.,  2007,
MNRAS, 374, 1310

\bibitem[\protect\citeauthoryear{{Chluba} \& {Sunyaev}}{{Chluba} \&
  {Sunyaev}}{2008b}]{Chluba_Sunyaev08b}
{Chluba} J.,  {Sunyaev} R.~A.,  2008b, A\&A, 480, 629

\bibitem[\protect\citeauthoryear{{Chluba} \& {Sunyaev}}{{Chluba} \&
  {Sunyaev}}{2009a}]{Chluba_Sunyaev09b}
{Chluba} J.,  {Sunyaev} R.~A.,  2009a, A\&A, 503, 345

\bibitem[\protect\citeauthoryear{{Chluba} \& {Sunyaev}}{{Chluba} \&
  {Sunyaev}}{2009b}]{Chluba_Sunyaev09a}
{Chluba} J.,  {Sunyaev} R.~A.,  2009b, A\&A, 496, 619

\bibitem[\protect\citeauthoryear{{Chluba} \& {Sunyaev}}{{Chluba} \&
  {Sunyaev}}{2010a}]{Chluba_Sunyaev10a}
{Chluba} J.,  {Sunyaev} R.~A.,  2010a, MNRAS, 402, 1221

\bibitem[\protect\citeauthoryear{{Chluba} \& {Sunyaev}}{{Chluba} \&
  {Sunyaev}}{2010b}]{Chluba_Sunyaev10b}
{Chluba} J.,  {Sunyaev} R.~A.,  2010b, A\&A, 512, A53+

\bibitem[\protect\citeauthoryear{{Chluba} \& {Thomas}}{{Chluba} \&
  {Thomas}}{2010}]{Chluba_Thomas10}
{Chluba} J., {Thomas} R.~M., 2010, MNRAS, pp 1876--+

\bibitem[\protect\citeauthoryear{{Chluba}, {Vasil} \& {Dursi}}{{Chluba}
  et~al.}{2010}]{Chluba_et_al10}
{Chluba} J.,  {Vasil} G.~M., {Dursi} L.~J.,  2010, MNRAS, 407, 599

\bibitem[\protect\citeauthoryear{{Dalgarno} \& {Lepp}}{{Dalgarno} \&
  {Lepp}}{1987}]{Dalgarno_Lepp87}
{Dalgarno} A., {Lepp} S., 1987, {Astrochemistry, IAU Symposium}, 120, 109

\bibitem[\protect\citeauthoryear{{Doroshkevich}, {Zeldovich} \&
  {Novikov}}{{Doroshkevich} et~al.}{1967}]{Doroshkevich_et_al67}
{Doroshkevich} A.~G.,  {Zeldovich} Y.~B., {Novikov} I.~D.,  1967, Soviet
  Astronomy, 11, 233

\bibitem[\protect\citeauthoryear{{Dubrovich}}{{Dubrovich}}{1975}]{Dubrovich75}
{Dubrovich} V.~K.,  1975, Soviet Astronomy Letters, 1, 196

\bibitem[\protect\citeauthoryear{{Dubrovich} \& {Grachev}}{{Dubrovich} \&
  {Grachev}}{2004}]{Dubrovich_Grachev04}
{Dubrovich} V.~K.,  {Grachev} S.~I.,  2004, Astronomy Letters, 30, 657

\bibitem[\protect\citeauthoryear{{Dubrovich} \& {Grachev}}{{Dubrovich} \&
  {Grachev}}{2005}]{Dubrovich_Grachev05}
{Dubrovich} V.~K.,  {Grachev} S.~I.,  2005, Astronomy Letters, 31, 359

\bibitem[\protect\citeauthoryear{{Dubrovich}, {Grachev} \&
  {Romanyuk}}{{Dubrovich} et~al.}{2009}]{Dubrovich_et_al09}
{Dubrovich} V.~K.,  {Grachev} S.~I.,    {Romanyuk} V.~G.,  2009, Astronomy
  Letters, 35, 723

\bibitem[\protect\citeauthoryear{{Dubrovich} \& {Stolyarov}}{{Dubrovich} \&
  {Stolyarov}}{1995}]{Dubrovich_Stolyarov95}
{Dubrovich} V.~K.,  {Stolyarov} V.~A.,  1995, A\&A, 302, 635

\bibitem[\protect\citeauthoryear{{Dubrovich} \& {Stolyarov}}{{Dubrovich} \&
  {Stolyarov}}{1997}]{Dubrovich_Stolyarov97}
{Dubrovich} V.~K.,  {Stolyarov} V.~A.,  1997, Astronomy Letters, 23, 565

\bibitem[\protect\citeauthoryear{{Fahr} \& {Loch}}{{Fahr} \&
  {Loch}}{1991}]{Fahr_Loch91}
{Fahr} H.~J.,  {Loch} R.,  1991, A\&A, 246, 1

\bibitem[\protect\citeauthoryear{{Fendt}, {Chluba}, {Rubi{\~n}o-Mart{\'{\i}}n}
  \& {Wandelt}}{{Fendt} et~al.}{2009}]{Fendt_et_al09}
{Fendt} W.~A.,  {Chluba} J.,  {Rubi{\~n}o-Mart{\'{\i}}n} J.~A., {Wandelt}
B.~D.,  2009, ApJS, 181, 627

\bibitem[\protect\citeauthoryear{{Goldman}}{{Goldman}}{1989}]{Goldman89}
{Goldman} S.~P., 1989, Phys. Rev. A, 40, 1185 

\bibitem[\protect\citeauthoryear{{Grachev}}{{Grachev}}{1988}]{Grachev88}
{Grachev} S.~I., 1988, Astrophysics, 28, 119 

\bibitem[\protect\citeauthoryear{{Grachev} \& {Dubrovich}}{{Grachev} \&
  {Dubrovich}}{1991}]{Grachev_Dubrovich91}
{Grachev} S.~I.,  {Dubrovich} V.~K.,  1991, Astrophysics, 34, 124

\bibitem[\protect\citeauthoryear{{Grachev} \& {Dubrovich}}{{Grachev} \&
  {Dubrovich}}{2008}]{Grachev_Dubrovich08}
{Grachev} S.~I.,  {Dubrovich} V.~K.,  2008, Astronomy Letters, 34, 439

\bibitem[\protect\citeauthoryear{{Grachev} \& {Dubrovich}}{{Grachev} \&
  {Dubrovich}}{2010}]{Grachev_Dubrovich10}
{Grachev} S.~I.,  {Dubrovich} V.~K.,  2010, arxiv:1010.4455

\bibitem[\protect\citeauthoryear{{Grin} \& {Hirata}}{{Grin} \&
  {Hirata}}{2010}]{Grin_Hirata10}
{Grin} D.,  {Hirata} C.~M.,  2010, Phys. Rev. D, 81, 083005

\bibitem[\protect\citeauthoryear{{Hirata}}{{Hirata}}{2008}]{Hirata08}
{Hirata} C.~M.,  2008, Phys. Rev. D, 78, 023001

\bibitem[\protect\citeauthoryear{{Hirata} \& {Forbes}}{{Hirata} \&
  {Forbes}}{2009}]{Hirata_Forbes09}
{Hirata} C.~M.,  {Forbes} J.,  2009, Phys. Rev. D, 80, 023001

\bibitem[\protect\citeauthoryear{{Hirata} \& {Switzer}}{{Hirata} \&
  {Switzer}}{2008}]{Hirata_Switzer08}
{Hirata} C.~M.,  {Switzer} E.~R.,  2008, Phys. Rev. D, 77, 083007

\bibitem[\protect\citeauthoryear{{Hu}, {Scott}, {Sugiyama} \& {White}}{{Hu}
  et~al.}{1995}]{Hu_et_al95}
{Hu} W.,  {Scott} D.,  {Sugiyama} N.,    {White} M.,  1995, Phys. Rev. D, 52,
  5498

\bibitem[\protect\citeauthoryear{{Hummer} \& {Storey}}{{Hummer} \&
  {Storey}}{1998}]{Hummer_Storey98}
{Hummer} D.~G.,  {Storey} P.~J.,  1998, MNRAS, 297, 1073

\bibitem[\protect\citeauthoryear{{Jones} \& {Wyse}}{{Jones} \&
  {Wyse}}{1985}]{Jones_Wyse85}
{Jones} B.~J.~T.,  {Wyse} R.~F.~G.,  1985, A\&A, 149, 144

\bibitem[\protect\citeauthoryear{{Kholupenko} \& {Ivanchik}}{{Kholupenko} \&
  {Ivanchik}}{2006}]{Kholupenko_Ivanchik06}
{Kholupenko} E.~E.,  {Ivanchik} A.~V.,  2006, Astronomy Letters, 32, 795

\bibitem[\protect\citeauthoryear{{Kholupenko}, {Ivanchik} \&
  {Varshalovich}}{{Kholupenko} et~al.}{2007}]{Kholupenko_et_al07}
{Kholupenko} E.~E.,  {Ivanchik} A.~V.,    {Varshalovich} D.~A.,  2007, MNRAS,
  378, L39

\bibitem[\protect\citeauthoryear{{Kholupenko}, {Ivanchik} \&
  {Varshalovich}}{{Kholupenko} et~al.}{2008}]{Kholupenko_et_al08}
{Kholupenko} E.~E.,  {Ivanchik} A.~V.,    {Varshalovich} D.~A.,  2008,
  Astronomy Letters, 34, 725

\bibitem[\protect\citeauthoryear{{Kholupenko}, {Ivanchik} \&
  {Varshalovich}}{{Kholupenko} et~al.}{2010}]{Kholupenko_et_al10}
{Kholupenko} E.~E.,  {Ivanchik} A.~V.,    {Varshalovich} D.~A.,  2010, Phys.
  Rev. D, 81, 083004

\bibitem[\protect\citeauthoryear{{Labzowsky}, {Solovyev} \&
  {Plunien}}{{Labzowsky} et~al.}{2009}]{Labzowsky_et_al09}
{Labzowsky} L.,  {Solovyev} D., {Plunien} G.,  2009, 
Phys. Rev. A, 80, 062514 

\bibitem[\protect\citeauthoryear{{Lyubarsky} \& {Sunyaev}}{{Lyubarsky} \&
  {Sunyaev}}{1983}]{Lyubarsky_Sunyaev83}
{Lyubarsky} Y.~E.,  {Sunyaev} R.~A.,  1983, A\&A, 123, 171

\bibitem[\protect\citeauthoryear{{Ma} \& {Bertschinger}}{{Ma} \&
  {Bertschinger}}{1995}]{Ma_Bertschinger95}
{Ma} C.,  {Bertschinger} E.,  1995, ApJ, 455, 7

\bibitem[\protect\citeauthoryear{{Matsuda}, {Sat{\= o}} \& {Takeda}}{{Matsuda}
  et~al.}{1969}]{Matsuda_et_al69}
{Matsuda} T.,  {Sat{\= o}} H.,    {Takeda} H.,  1969, Progress of Theoretical
  Physics, 42, 219

\bibitem[\protect\citeauthoryear{{Nussbaumer} \& {Schmutz}}{{Nussbaumer} \&
  {Schmutz}}{1984}]{Nussbaumer_Schmutz84}
{Nussbaumer} H.,  {Schmutz} W.,  1984, A\&A, 138, 495

\bibitem[\protect\citeauthoryear{{Peebles}}{{Peebles}}{1965}]{Peebles65}
{Peebles} P.~J.~E.,  1965, ApJ, 142, 1317

\bibitem[\protect\citeauthoryear{{Peebles}}{{Peebles}}{1968}]{Peebles68}
{Peebles} P.~J.~E.,  1968, ApJ, 153, 1

\bibitem[\protect\citeauthoryear{{Pequignot}, {Petitjean} \&
  {Boisson}}{{Pequignot} et~al.}{1991}]{Pequignot_et_al91}
{Pequignot} D.,  {Petitjean} P.,    {Boisson} C.,  1991, A\&A, 251, 680

\bibitem[\protect\citeauthoryear{{Rubi{\~n}o-Mart{\'{\i}}n}, {Chluba} \&
  {Sunyaev}}{{Rubi{\~n}o-Mart{\'{\i}}n} et~al.}{2008}]{Rubino_Martin_et_al08}
{Rubi{\~n}o-Mart{\'{\i}}n} J.~A.,  {Chluba} J.,    {Sunyaev} R.~A.,  2008,
  A\&A, 485, 377

\bibitem[\protect\citeauthoryear{{Rybicki} \& {dell'Antonio}}{{Rybicki} \&
  {dell'Antonio}}{1993}]{Rybicki_dell_Antonio93}
{Rybicki} G.~B.,  {dell'Antonio} I.~P.,  1993, Observational Cosmology,
  Astronomical Society of the Pacific Conference Series, 51, 548

\bibitem[\protect\citeauthoryear{{Sc{\'o}ccola}, {Landau} \&
  {Vucetich}}{{Sc{\'o}ccola} et~al.}{2008}]{Scoccola_et_al08}
{Sc{\'o}ccola} C.~G.,  {Landau} S.~J.,    {Vucetich} H.,  2008, Physics Letters
  B, 669, 212

\bibitem[\protect\citeauthoryear{{Scott} \& {Moss}}{{Scott} \&
  {Moss}}{2009}]{Scott_Moss09}
{Scott} D.,  {Moss} A.,  2009, MNRAS, 397, 445

\bibitem[\protect\citeauthoryear{{Seager}, {Sasselov} \& {Scott}}{{Seager}
  et~al.}{1999}]{Seager_et_al99}
{Seager} S.,  {Sasselov} D.~D.,    {Scott} D.,  1999, ApJ, 523, L1

\bibitem[\protect\citeauthoryear{{Seager}, {Sasselov} \& {Scott}}{{Seager}
  et~al.}{2000}]{Seager_et_al00}
{Seager} S.,  {Sasselov} D.~D.,    {Scott} D.,  2000, ApJS, 128, 407

\bibitem[\protect\citeauthoryear{{Shapiro} \& {Breit}}{{Shapiro} \&
  {Breit}}{1959}]{Shapiro_Breit59}
{Shapiro} J.,  {Breit} G.,  1959, Phys. Rev., 113, 179

\bibitem[\protect\citeauthoryear{{Shaw} \& {Chluba}}{{Shaw} \&
  {Chluba}}{2011}]{Shaw_Chluba11}
{Shaw} J.~R.,  {Chluba} J.,  2011, arxiv:1102.3683

\bibitem[\protect\citeauthoryear{{Sunyaev} \& {Chluba}}{{Sunyaev} \&
  {Chluba}}{2009}]{Sunyaev_Chluba09}
{Sunyaev} R.~A.,  {Chluba} J.,  2009, Astronomische Nachrichten, 330, 657

\bibitem[\protect\citeauthoryear{{Sunyaev} \& {Zeldovich}}{{Sunyaev} \&
  {Zeldovich}}{1970}]{Sunyaev_Zeldovich70}
{Sunyaev} R.~A.,  {Zeldovich} Y.~B.,  1970, Astrophysics and Space Science, 7, 3

\bibitem[\protect\citeauthoryear{{Switzer} \& {Hirata}}{{Switzer} \&
  {Hirata}}{2008a}]{Switzer_Hirata08a}
{Switzer} E.~R.,  {Hirata} C.~M.,  2008a, Phys. Rev. D, 77, 083006

\bibitem[\protect\citeauthoryear{{Switzer} \& {Hirata}}{{Switzer} \&
  {Hirata}}{2008b}]{Switzer_Hirata08b}
{Switzer} E.~R.,  {Hirata} C.~M.,  2008b, Phys. Rev. D, 77, 083008

\bibitem[\protect\citeauthoryear{{Verner} \& {Ferland}}{{Verner} \&
  {Ferland}}{1996}]{Verner_Ferland96}
{Verner} D.~A.,  {Ferland} G.~J.,  1996, ApJS, 103, 467

\bibitem[\protect\citeauthoryear{{Wong} \& {Scott}}{{Wong} \& {Scott}}
{2008}]{Wong_Scott07}
{Wong} W.~Y.,  {Scott} D.,  2007, MNRAS, 375, 1441

\bibitem[\protect\citeauthoryear{{Wong}, {Moss} \& {Scott}}{{Wong}
  et~al.}{2008}]{Wong_et_al08}
{Wong} W.~Y.,  {Moss} A.,    {Scott} D.,  2008, MNRAS, 386, 1023

\bibitem[\protect\citeauthoryear{{Zeldovich}, {Kurt} \& {Syunyaev}}{{Zeldovich}
  et~al.}{1968}]{Zeldovich_et_al68}
{Zeldovich} Y.~B.,  {Kurt} V.~G.,    {Syunyaev} R.~A.,  1968, Zhurnal
  Eksperimentalnoi i Teoreticheskoi Fiziki, 55, 278

\bibitem[\protect\citeauthoryear{{Zon} \& {Rapoport}}{{Zon} \&
  {Rapoport}}{1968}]{Zon_Rapoport68}
{Zon} B.~A.,  {Rapoport} L.~P.,  1968, Soviet Journal of Experimental and
  Theoretical Physics Letters, 7, 52
\end{thebibliography}
\end{document}